\documentclass[preprint,showpacs,preprintnumbers,amssymb,aps,superscriptaddress]{revtex4}
\usepackage{graphicx}% Include figure files
\usepackage{dcolumn}% Align table columns on decimal point
\usepackage{bm}% bold math

\begin{document}
\title{Rigorous criterion for reentrance in the spin-1/2 Ising-Heisenberg model on diamond-like decorated Bethe lattices}
\author{J. STRE\v{C}KA} 
\affiliation{Department of Theoretical Physics and Astrophysics, Faculty of Science, \\ 
P. J. \v{S}af\'{a}rik University, Park Angelinum 9, 040 01 Ko\v{s}ice, Slovak Republic} 
\author{C. EKIZ} 
\affiliation{Department of Physics, Faculty of Science, Adnan Menderes University, \\
090 10 Aydin, Turkey}
      
\begin{abstract}
The spin-1/2 Ising-Heisenberg model on diamond-like decorated Bethe lattices is exactly solved 
with the help of decoration-iteration transformation and exact recursion relations. 
It is shown that the model under investigation exhibits reentrant phase transitions whenever 
a sufficiently high coordination number of the underlying Bethe lattice is considered.
\end{abstract}

\maketitle

\section{Introduction}
The ferromagnetic spin-1/2 Ising-Heisenberg model on planar lattices currently attracts a great deal 
of attention as it may exhibit a variety of interesting phenomena such as a quantum critical point \cite{stre02,stre06}, peculiar spontaneous antiferromagnetic long-range order \cite{stre02} or several unusual spin-liquid ground states \cite{stre06}. In this work, the spin-1/2 Ising-Heisenberg model on diamond-like decorated Bethe lattices will be treated exactly with the aim to study its ground-state and finite-temperature phase diagrams. 

\section{Ising-Heisenberg model and its exact solution}

\begin{figure}[t]
\begin{center}
\includegraphics[width=8.5cm]{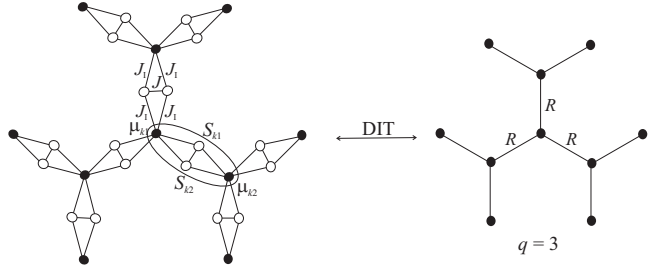} \\
\end{center}
\vspace{-0.6cm}
\caption{A schematic representation of the decoration-iteration transformation (DIT) between the spin-1/2 Ising-Heisenberg model on the diamond-like decorated Bethe lattice (figure on the left) 
and the spin-1/2 Ising model on the simple Bethe lattice with the coordination number $q=3$ 
(figure on the right). The full (empty) circles denote lattice positions of the Ising (Heisenberg) spins and the ellipse demarcates all interaction terms belonging to the $k$th bond 
Hamiltonian (\ref{eq:2}).}
\label{fig:1}
\end{figure}

Consider the spin-1/2 Ising-Heisenberg model on the diamond-like decorated Bethe lattices
schematically illustrated on the left-hand-side of Fig.~\ref{fig:1}. In this figure, 
the full circles denote lattice positions of the Ising spins $\mu$, the empty circles 
label lattice positions of the Heisenberg spins $S$ and the parameter $q$ is the 
coordination number of the underlying Bethe lattice. For further convenience, 
the total Hamiltonian can be defined as a sum of the bond Hamiltonians, 
i.e. $\hat{\cal H} = \sum_{k} \hat{\cal H}_k$, where each bond Hamiltonian $\hat{\cal H}_k$ 
involves all the interaction terms belonging to the $k$th diamond unit (see Fig.~\ref{fig:1}) 
\begin{eqnarray}
\hat{\cal H}_{k} = \!\!\! &-& \!\!\! J [\Delta (\hat S_{k1}^x \hat S_{k2}^x 
   + \hat S_{k1}^y \hat S_{k2}^y) + \hat S_{k1}^z \hat S_{k2}^z ] \nonumber \\
   \!\!\! &-& \!\!\! J_{\rm I} (\hat S_{k1}^z + \hat S_{k2}^z)(\hat \mu_{k1}^z + \hat \mu_{k2}^z)
\label{eq:2}
\end{eqnarray}
In above, $\hat S_i^{\alpha}$ ($\alpha = x,y,z$) and $\hat \mu_j^z$ label spatial components 
of the spin-1/2 operator, the former term $J$ accounts for the XXZ interaction between 
the nearest-neighbour Heisenberg spins, $\Delta$ is a spatial anisotropy in this interaction 
and the latter term $J_{\rm I}$ accounts for the Ising-type interaction between the nearest-neighbour 
Ising and Heisenberg spins, respectively.  

A crucial step of our approach lies an evaluation of the partition function. Due to a commutability 
of different bond Hamiltonians, the partition function can be partially factorized to a product 
of bond partition functions 
\begin{equation}
{\cal Z}_{\rm IHM} = \sum_{\{ \mu_i \}} \prod_{k = 1}^{Nq/2} \mbox{Tr}_k \exp(- \beta \hat {\cal H}_k)
               = \sum_{\{ \mu_i \}} \prod_{k = 1}^{Nq/2} {\cal Z}_{k},
\label{eq:3}
\end{equation}
where $\beta = 1/(k_{\rm B} T)$, $k_{\rm B}$ is Boltzmann's constant, $T$ denotes absolute temperature,
$N$ is the total number of the Ising spins, the symbol $\mbox{Tr}_k$ stands for a trace over degrees 
of freedom of the $k$th Heisenberg spin pair and the summation $\sum_{\{ \mu_i \}}$ runs over all available configurations of the Ising spins. A calculation of the bond partition function 
${\cal Z}_{k}$ can easily be accomplished through a direct diagonalization of the bond Hamiltonian (\ref{eq:2}). Subsequently, the gained expression for ${\cal Z}_{k}$ will depend just on two outer Ising spins and can be replaced with the generalized decoration-iteration transformation \cite{stre06}
\begin{eqnarray}
{\cal Z}_k \!\!\! && \!\!\! (\mu_{k1}^z, \mu_{k2}^z) = 2 \exp(\beta J/4) \cosh[\beta J_{\rm I}(\mu_{k1}^z + \mu_{k2}^z)] + \nonumber \\ 
\!\!\!\! && \!\!\!\! 2 \exp(-\beta J/4) \cosh (\beta J \Delta/2) = 
A \exp(\beta R \mu_{k1}^z \mu_{k2}^z). 
\label{eq:5}
\end{eqnarray}
The mapping parameters $A$ and $R$ can directly be obtained from the decoration-iteration 
transformation (\ref{eq:5}), which holds quite generally if and only if  
\begin{eqnarray}
A \!\! &=& \!\!  [{\cal Z}_k (1/2, 1/2) {\cal Z}_k (1/2, -1/2)]^{1/2},  \label{eq:6} \\
\beta R \!\! &=& \!\! 2 \ln [{\cal Z}_k (1/2, 1/2)] - 2 \ln [{\cal Z}_k (1/2, -1/2)].  
\label{eq:7}
\end{eqnarray}
Substituting the algebraic transformation (\ref{eq:5}) into the formula (\ref{eq:3}) one gets a precise mapping relationship between partition functions of the spin-1/2 Ising-Heisenberg model on the diamond-like decorated Bethe lattice and the corresponding spin-1/2 Ising model on the simple Bethe lattice (see Fig.~\ref{fig:1})
\begin{eqnarray}
{\cal Z}_{\rm IHM} (\beta, J, \Delta, J_{\rm I}) = A^{Nq/2} {\cal Z}_{\rm IM} (\beta, R).
\label{eq:11} 
\end{eqnarray}
Note that the partition function of the spin-1/2 Ising model on the Bethe lattice can exactly be calculated by making use of exact recursion relations \cite{baxt82} and hence, our exact calculation is 
formally completed. Using the mapping equivalence (\ref{eq:11}) between both partition functions, 
the sublattice magnetization of the Ising spins directly equals to the single-site magnetization 
of the corresponding spin-1/2 Ising model on the simple Bethe lattice  
\begin{eqnarray}
m_{i}^{z} \equiv \langle \hat \mu_{k}^z \rangle_{\rm IHM} 
= \langle \hat \mu_{k}^z \rangle_{\rm IM} \equiv m_{\rm IM} (\beta R).
\label{eq:12} 
\end{eqnarray}
Above, the symbols $\langle \cdots \rangle_{\rm IHM}$ and $\langle \cdots \rangle_{\rm IM}$ mark 
canonical ensemble average performed within two models connected through the mapping relation (\ref{eq:11}). Note furthermore that the corresponding exact result for the magnetization $m_{\rm IM}$ can be also found with the help of recursion relations \cite{baxt82}. The sublattice magnetization of the Heisenberg 
spins follows from the formula 
\begin{eqnarray}
m_{h}^{z} \equiv \langle \hat S_{k}^z \rangle_{\rm IHM} =  \frac{m_{\rm IM} \sinh(\beta J_{\rm I})} {\cosh(\beta J_{\rm I}) + \exp(-\frac{\beta J}{2}) \cosh \Bigl(\frac{\beta J \Delta}{2} \Bigr)}.
\label{eq:13} 
\end{eqnarray}

\section{Results and Discussion}
\label{sec:result}

First, let us take a closer look at the ground state of the spin-1/2 Ising-Heisenberg model with both ferromagnetic interactions ($J>0$, $J_{\rm I}>0$). The classical ferromagnetic phase 
with all Ising as well as Heisenberg spins aligned parallel constitutes the ground state whenever 
the exchange anisotropy $\Delta < \Delta_c = 1 + 2J_{\rm I}/J$. In the opposite case $\Delta > \Delta_c$, the remarkable spin-liquid phase constitutes the ground state and this disordered phase 
can be characterized through a complete randomness of the Ising spins and the quantum entanglement of 
Heisenberg spin pairs. An origin of this unusual disordered phase lies in the geometric 
frustration, which is triggered by a competition between the easy-axis Ising interaction 
and the easy-plane XXZ Heisenberg interaction \cite{stre06}.

\begin{figure}[htb]
\vspace{0.3cm}
\begin{center}
\includegraphics[width=6.4cm]{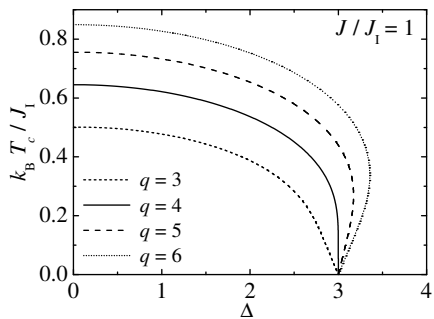} 
\end{center}
\vspace{-1.2cm}
\caption{Critical temperature as a function of the exchange anisotropy $\Delta$ for one 
selected ratio $J/J_{\rm I} = 1$ and several values of the coordination number $q$.}
\label{fig:2}
\end{figure}
\begin{figure}[htb]
\vspace{-0.3cm}
\begin{center}
\includegraphics[width=6.5cm]{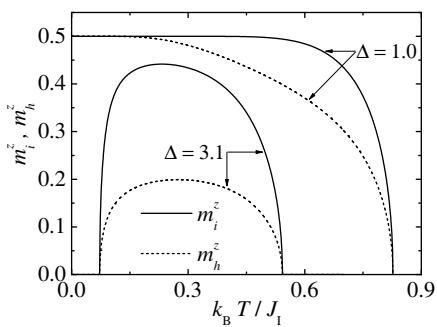} 
\end{center}
\vspace{-1.2cm}
\caption{Thermal variations of both spontaneous sublattice magnetizations for $J/J_{\rm I} = 1$, 
$q=6$ and two different values of the exchange anisotropy $\Delta$.}
\label{fig:3}
\end{figure}

Fig.~\ref{fig:2} shows the critical temperature as a function of the exchange anisotropy $\Delta$ 
for one particular ratio $J/J_{\rm I} = 1$ and several values of the coordination number $q$.
As one can see, the critical temperature vs. exchange anisotropy dependence basically changes 
with the coordination number $q$ of the underlying Bethe lattice. The critical lines are approaching the zero-temperature phase boundary between the classical ferromagnetic phase and disordered phase 
at $\Delta_c$ with a negative (positive) slope for the diamond-like decorated Bethe lattices 
with the coordination number $q<4$ ($q>4$) and with an infinite gradient for the particular case 
with the coordination number $q=4$. Owing to this fact, reentrant phase transitions can be observed 
for $\Delta \gtrsim \Delta_c$ only when assuming the diamond-like decorated Bethe lattices 
with a sufficiently high coordination number $q>4$. Thermal variations of both sublattice magnetizations, which are depicted in Fig.~\ref{fig:3}, provide an independent confirmation 
of the aforedescribed reentrance. 

In conclusion, it is worthy to note that the rigorous procedure developed on the grounds of decoration-iteration transformation and exact recursion relations can be readily adapted to 
treat the investigated model in a presence of external magnetic field, which will be examined 
in detail in our forthcoming work.

{\bf Acknowledgments}: This work was supported under the grants VEGA 1/0431/10 and VVGS 2/09-10.

\end{document}